\documentstyle [11pt]{article}
\def\ZzZ{{\hbox{\tenrm Z\kern-.31em{Z}}}}
\def\CcC{{\hbox{\tenrm C\kern-.45em{\vrule height.67em width0.08em depth-
.04em \hskip.45em }}}}

\def\mapbelow#1{\smash{\mathop{\longrightarrow}\limits_{#1}}}
\newcommand{\lab}{\label}
\newcommand{\non}{\nonumber}

\newcommand{\bc}{\begin{center}}
\newcommand{\ec}{\end{center}}
\newcommand{\be}{\begin{equation}}
\newcommand{\ee}{\end{equation}}
\newcommand{\bea}{\begin{eqnarray}}
\newcommand{\eea}{\end{eqnarray}}
\newcommand{\bs}{\begin{subequations}}
\newcommand{\es}{\end{subequations}}
\newcommand{\beq}{\begin{eqalignno}}
\newcommand{\eeq}{\end{eqalignno}}

\def\lab{\label}

\def\lf{\left}

\def\non{\nonumber}

\def\ri{\right}

\def\ka{\kappa}

\def\lab{\label}

\begin{document}

\bc {

{\bf Quantum noise induced entanglement and chaos}

{\bf  in the dissipative quantum model of brain}

\bigskip
\bigskip

Eliano Pessa$^\dag$ and Giuseppe Vitiello$^{\dag\dag}$

\medskip

$^{\dag}$Dipartimento di Psicologia, Universit\`a di Pavia, 27100
Pavia, Italy

$^{\dagger\dagger}$Dipartimento di Fisica ``E.R.Caianiello",
Universit\`a di Salerno, 84100 Salerno, Italy

INFN, Gruppo Collegato di Salerno and INFM, Sezione di Salerno

e-mail: pessa@unipv.it,  vitiello@sa.infn.it }

%$$ $$
\bigskip

\ec

{\bf Abstract} We discuss some features of the dissipative quantum
model of brain in the frame of the formalism of quantum
dissipation. Such a formalism is based on the doubling of the
system degrees of freedom. We show that the doubled modes account
for the quantum noise in the fluctuating random force in the
system-environment coupling. Remarkably, such a noise manifests
itself through the coherent structure of the system ground state.
The entanglement of the system modes with the doubled modes is
shown to be permanent in the infinite volume limit. In such a
limit the trajectories in the memory space are classical chaotic
trajectories.

\smallskip

%PACS: ******

\bigskip
\bigskip

\section{Introduction}
%
%\vspace{.3cm}

In this paper we discuss chaos, entanglement and quantum noise in
the dissipative quantum model of brain
\cite{Vitiello:1995wv,MyD,Alfinito:2000ck,Pessa:1999fp}. This
model is an extension to dissipative dynamics of the quantum field
theory (QFT) model of brain originally formulated by Umezawa and
Ricciardi in 1967 \cite{UR} and subsequently developed by Stuart,
Takahashi and Umezawa \cite{STU}, by Jibu and Yasue \cite{JY} and
by Jibu, Pribram and Yasue \cite{JPY}. For a general account of
the model see ref \cite{MyD}.

As we will see from the outcomes of our discussion, we will be
lead to uncover some consequences of this model which seem to be
related, although the precise relation has not yet been worked out
in detail, to actual experimental findings in neurobiology
\cite{Free}.

The time scales of the working brain are such that it is extremely
hard, if not impossible, to think that they can be obtained in a
classical approach: the configuration space of the working brain
is so large that only a quantum dynamics may account of the very
short time intervals needed to the brain to span it in such an
efficient way as it does.

In the QFT model the memory storage is described in terms of the
coherent Bose condensation process in the system lowest energy
state (usually called the ground state or else the vacuum state).
Bose condensation occurs as a consequence of the action on the
brain of the external inputs. These break the symmetry of the
quantum field dynamics. The quantum fields are the dipole
vibrational quantum fields associated with the water and other
bio-molecules endowed with static and/or radiative electric dipole
moment.

According to the Goldstone theorem in QFT \cite{ITZ,umezawa}, the
spontaneous breakdown of the symmetry implies the existence of
long-range correlation modes (also called the Nambu-Goldstone (NG)
modes) in the ground state of the system. These modes are massless
modes in the infinite volume limit, but they may acquire a finite,
non-zero mass due to boundary or impurity effects
\cite{Alfinito:2001mm}. In the quantum model of brain these modes
are called dipole-wave-quanta (DWQ). The density of their
condensation in the ground states acts as a {\it code} classifying
the state and the memory there recorded. States with different
code values are unitarily inequivalent states, i.e there is no
unitary transformation relating states of different codes.

In formulating a mathematical model of brain we cannot avoid to
take into consideration the dissipative character of its dynamics,
since the brain is an intrinsically open system, continuously
interacting with the environment. As elsewhere observed
\cite{Vitiello:1995wv,MyD}, the very same fact of ``getting an
information" introduces a partition in the time coordinate, so
that one may distinguish between {\it before} ``getting the
information" (the past) and {\it after} ``getting the information"
(the future): the {\it arrow of time} is in this way introduced.
...``{\it Now} you know it!" is the familiar warning to mean that
now, i.e. after having received a certain information, you are not
the same person as before getting it. It has been shown
\cite{Alfinito:2000ck} that the psychological arrow of time
(arising as an effect of memory recording) points in the same
direction of the thermodynamical arrow of time (increasing entropy
direction) and of the cosmological arrow of time (the expanding
Universe direction) \cite{Alfinito:1999bv}.

The canonical quantization procedure of a dissipative system
requires to include in the formalism also the system representing
the environment (or heat bath) in which the system is embedded.
One possible way to do that is to depict the environment as the
time-reversal image of the system \cite{Celeghini:yv}: the
environment is thus described as the {\it Double} of the system in
the time-reversed dynamics (the system image in the mirror of
time).

Of course, the specific details of the system--environment
coupling may be very intricate and changeable so that they are
difficult to be measured and known. One possibility is to take
into account the environmental influence on the brain by a
suitable {\it choice} of the brain vacuum state among the
infinitely many of them. Such a choice is triggered by the
external input (breakdown of the symmetry), and it actually is the
end point of the internal (spontaneous) dynamical process of the
brain (self-organization). The chosen vacuum thus carries the {\it
signature} (memory) of the reciprocal brain--environment influence
at a given time under given boundary conditions. A change in the
brain--environment reciprocal influence then would correspond to a
change in the choice of the brain vacuum: the brain evolution
through the vacuum states is thus the evolution of the coupling of
the brain with the surrounding world.

Within this paper we will adopt, from the starting, the
mathematical framework of QFT. This implies that the brain system
will be described in terms of an infinite collection of damped
harmonic oscillators $A_{\kappa}$ (the simplest prototype of a
dissipative system) representing the DWQ \cite{Vitiello:1995wv}.

The collection of damped oscillators is ruled by the Hamiltonian
\cite{Vitiello:1995wv,Celeghini:yv}
\be\lab{1}
 H = H_{0}   + H_{I} ,
\ee
\be\lab{2}
 H_{0} = \sum_{\kappa} \hbar \Omega_{\kappa}  \bigl (
A_{\kappa}^{\dagger} A_{\kappa} - {\tilde A}_{\kappa}^{\dagger}
{\tilde A}_{\kappa} \bigr )~,~~~~ H_{I} = i \sum_{\kappa} \hbar
\Gamma_{\kappa} \bigl ( A_{\kappa}^{\dagger} {\tilde
A}_{\kappa}^{\dagger} - A_{\kappa} {\tilde A}_{\kappa} \bigr )~,
\ee
where $\Omega_{\kappa}$ is the frequency and $\Gamma_{\kappa}$ is
the damping constant. The ${\tilde A}_{\kappa}$ modes are the
"time-reversed mirror image" (the ``mirror modes'') of the
$A_{\kappa}$ modes. They are the doubled modes and represent the
environment modes. $\kappa$ generically labels the mode degrees of
freedom, {\it e.g.} spatial momentum (see
\cite{Vitiello:1995wv,Celeghini:yv} for details).

Since the environment is described in the quantum dissipation
formalism by the doubled degrees of freedom, from now on we will
use, without further specification, the word environment meaning
such a set of doubled degrees of freedom.

The paper is organized as follows. In Section 2 we show that the
doubled $\tilde A$ modes actually account for the quantum noise in
the fluctuating random force in the system-environment coupling.
Section 3 contains a discussion of the entanglement between
non--tilde and tilde mode sectors induced by quantum noise.
Section 4 is devoted to show that we may have chaotic trajectories
(in the sense of dynamical system theory) in the space of the
memory states, a result which also suggests a possible connection
with laboratory observations \cite{Free}. Concluding remarks are
presented in Section 5. An extended report of the results obtained
in this paper has been presented in \cite{MM}.

It has been suggested that the doubled degrees of freedom may play
some role in the discussion of consciousness mechanisms
\cite{Vitiello:1995wv}. However, we leave out of the present paper
the discussion on such a topic (the interested reader may consult
refs. \cite{Vitiello:1995wv,MyD} and the references there quoted).

\medskip

\section{Doubling and quantum noise}

As a preliminary to the discussion presented in the following
sections and in order to better understand the role played by the
quantum noise in the dissipative brain model, in this section we
shortly summarize the description of dissipative systems in the
frame of the quantum Brownian motion as described by Schwinger
\cite{Schw} and by Feynman and Vernon \cite{FeyVer}.

We essentially refer to the results derived in refs.
\cite{Srivastava:1995yf,Blasone:1997xt}, where the doubling of the
phase-space degrees of freedom is discussed and the doubled
variables are shown to account for the quantum noise effects in
the fluctuating random force in the system-environment coupling.
This result adds a new perspective to the doubling in the quantum
model of brain. It seem to point to a relation with some
experimental observations in the brain behavior \cite{Free}.
However, we leave to a future work the deeper analysis which is
needed in order to show the details of such a relation.

To be definite, we consider the damped harmonic oscillator (dho)
\be\lab{3} m \ddot{x}+\gamma\dot{x}+\ka x=0~, \ee
as a simple prototype for dissipative systems. However, our
results also apply to more general systems than the one
represented in (\ref{3}).

The damped oscillator Eq. (\ref{3}) is a non-hamiltonian system
and therefore the customary canonical quantization procedure
cannot be followed. However, one can face the problem by resorting
to well known tools such as the density matrix and the Wigner
function.

It is instructive to consider first the special case of a particle
in the absence of friction with Hamiltonian
\be\lab{H} H=- \frac{\hbar^2}{2m}\left(\frac{\partial}{\partial
x}\right)^2 +V(x)~. \ee
The density matrix equation of motion is given by
\be\lab{5} i\hbar \frac{d \rho}{dt}=[H,\rho ]~. \ee
The density matrix function is
\be\lab{8} \langle x+\frac{1}{2}y|\rho (t)|x-\frac{1}{2}y\rangle =
\psi^* \lf(x+\frac{1}{2}y,t\ri)\psi \lf(x-\frac{1}{2}y,t\ri)
\equiv W(x,y,t)~, \ee
with the associated standard expression for the Wigner function
\cite{Feynman,Haken},

\be\lab{W} W(p,x,t) = \frac{1}{2\pi \hbar}\int { W(x,y,t)
e^{\lf(-i\frac{py}{\hbar}\ri)}dy}~. \ee

In the coordinate representation, by introducing the notation
\be\lab{7} x_{\pm}=x\pm \frac{1}{2}y~, \ee
Eq. (\ref{5}) is written as
\bea \non && i\hbar \frac{\partial}{\partial t}\langle x_+|\rho
(t)|x_-\rangle=
\\ \lab{6}
&&\lf\{ -\frac{\hbar^2}{2m}\lf[\lf(\frac{\partial}{\partial
x_+}\ri)^2-\lf(\frac{\partial}{\partial x_-}\ri)^2\ri]
+[V(x_+)-V(x_-)] \ri\}\langle x_+|\rho (t)|x_-\rangle , \eea
and the equation for $W(p,x,t)$ is
\be\lab{9a} i\hbar \frac{\partial }{\partial t} W(x,y,t)={\cal
H}_o W(x,y,t) \ee
\be\lab{9b} {\cal H}_o=\frac{1}{m}p_xp_y
+V\lf(x+\frac{1}{2}y\ri)-V\lf(x-\frac{1}{2}y\ri), \ee
\be\lab{9bb} p_x=-i\hbar\frac{\partial}{\partial x},~~
p_y=-i\hbar\frac{\partial}{\partial y}. \ee
The Hamiltonian (\ref{9b}) may be obtained from the Lagrangian
\be\lab{10}
 {\cal L}_o=m
\dot{x}\dot{y}-V\lf(x+\frac{1}{2}y\ri)+V\lf(x-\frac{1}{2}y\ri).
\ee
We thus conclude that the density matrix and the Wigner function
formalism {\it requires}, even in the non-dissipative case (zero
mechanical resistance $\gamma$), the introduction of a ``doubled"
set of coordinates, $x_{\pm}$, or, alternatively, $x$ and $y$. One
may understand this as related to the introduction of the
``couple" of indices {\it necessary} to label the density matrix
elements (cf. Eq. (\ref{6})).

Let us now consider the case of the particle interacting with a
thermal bath at temperature $T$. Let $f$ denote the random force
on the particle at the position $x$ due to the bath. The
interaction Hamiltonian between the bath and the particle is
written as
\be\lab{11} H_{int}=-fx~. \ee

In the Feynman-Vernon formalism, the effective action for the
particle is given by
\be\lab{12} {\cal A}[x,y]=\int_{t_i}^{t_f}dt\,{\cal
L}_o(\dot{x},\dot{y},x,y) +{\cal I}[x,y], \ee
with ${\cal L}_o$ defined as in Eq.(\ref{10}) and
\be\lab{13} e^{\frac{i}{\hbar}{\cal I}[x,y]}\,=\, \langle  (
e^{-\frac{i}{\hbar}\int_{t_i}^{t_f}f(t)x_-(t)dt}\, )_-\, (
e^{\frac{i}{\hbar}\int_{t_i}^{t_f}f(t)x_+(t)dt} \,)_+\rangle. \ee
In Eq.(\ref{13}), the symbol $\langle ~.~ \rangle$ denotes average
with respect to the thermal bath; ``$(.)_{+}$'' and ``$(.)_{-}$''
denote time ordering and anti-time ordering, respectively; the
c-number coordinates $x_{\pm }$ are defined as in Eq.(\ref{7}). If
the interaction between the bath and the coordinate $x$ (i.e
$H_{int}=-fx$ ) were turned off, then the operator $f$ of the bath
would develop in time according to $f(t)=e^{iH_\gamma
t/\hbar}fe^{-iH_\gamma t/\hbar }$ where $H_\gamma$ is the
Hamiltonian of the isolated bath (decoupled from the coordinate
$x$). $f(t)$ is then the force operator of the bath to be used in
Eq.(\ref{13}).

The interaction ${\cal I}[x,y]$ between the bath and the particle
has been evaluated in ref. \cite{Srivastava:1995yf} for a linear
passive damping due to thermal bath by following Feynman and
Vernon \cite{FeyVer}, and Schwinger \cite{Schw}. The final result
is \cite{Srivastava:1995yf}:
\bea\nonumber {\cal
I}[x,y]&=&\frac{1}{2}\int_{t_i}^{t_f}dt\,[x(t)F_y^{ret}(t)+
y(t)F_x^{adv}(t)]
\\ \lab{29}
&&+\frac{i}{2\hbar}\int_{t_i}^{t_f}\int_{t_i}^{t_f}dt ds
\,N(t-s)y(t)y(s)~, \eea
where the retarded force on $y$ and the advanced force on $x$ are
given in terms of the retarded and advanced Greens functions
$G_{ret}(t-s)$ and $G_{adv}(t-s)$:
\be\lab{28a} F_y^{ret}(t)=\int_{t_i}^{t_f}ds \,G_{ret}(t-s)y(s),
~~~~ F_x^{adv}(t)=\int_{t_i}^{t_f}ds\, G_{adv}(t-s)x(s), \ee
respectively. In Eq (\ref{29}) $N(t-s)$ is the quantum noise in
the fluctuating random force and it is given by
\be\lab{21} N(t-s)=\frac{1}{2}\langle f(t)f(s)+f(s)f(t)\rangle.
\ee
The real and the imaginary part of the action are given by
\be\lab{30a} {\cal R}e{\cal A}[x,y]=\int_{t_i}^{t_f}dt\,{\cal L},
\ee
\be\lab{30b} {\cal L}\,=\,m \dot{x}\dot{y}-\lf[V(x+\frac{1}{2}
y)-V(x-\frac{1}{2}y)\ri] + \frac{1}{2}\lf[x F_y^{ret} + y
F_x^{adv}\ri], \ee
and
\be\lab{30c} {\cal I}m{\cal A}[x,y]=
\frac{1}{2\hbar}\int_{t_i}^{t_f}\int_{t_i}^{t_f}dt ds
\,N(t-s)y(t)y(s). \ee
respectively. Eqs.(\ref{30a}),~(\ref{30b}), and (\ref{30c}), are
{\it rigorously exact} results for linear passive damping due to
the bath.

They show that in the classical limit ``$\hbar \rightarrow 0$''
nonzero $y$ yields an ``unlikely process'' in view of the large
imaginary part of the action implicit in Eq.(\ref{30c}) (cf.
Eq.(\ref{29}) ). Nonzero $y$, indeed, may lead to a negative real
exponent in the evolution operator, which in the limit $\hbar
\rightarrow 0$ may produce a negligible contribution to the
probability amplitude. On the contrary, {\it at quantum level
nonzero $y$ accounts for quantum noise effects in the fluctuating
random force in the system-environment coupling arising from the
imaginary part of the action} \cite{Srivastava:1995yf}.

This is the conclusion we wanted to reach.

When in Eq.(\ref{30b}) we use $F_y^{ret}=\gamma\dot{y}$ and
$F_x^{adv}=-\gamma\dot{x}$, we get
\be\lab{2a} {\cal L}(\dot{x},\dot{y},x,y)\,=\,m \dot{x}\dot{y}
-V\lf(x+\frac{1}{2}y\ri)+V\lf(x-\frac{1}{2}y\ri)+\frac{\gamma}{2}(x\dot{y}-
y\dot{x})~. \ee
By using
\be\lab{P} V\lf(x \pm \frac{1}{2}y\ri) = \frac{1}{2}\ka (x \pm
\frac{1}{2}y)^{2}
\ee
in  Eq. (\ref{2a}), the dho equation (\ref{3}) and its
complementary equation for the $y$ coordinate
\be\lab{3cy} m \ddot{y}-\gamma\dot{y}+\ka y = 0 . \ee
are derived. The $y$-oscillator is the time--reversed image of the
$x$-oscillator (\ref{3}).

From the manifold of solutions to Eqs. (\ref{3}),~(\ref{3cy}) we
could choose those for which the $y$ coordinate is constrained to
be zero, then Eqs. (\ref{3}) and (\ref{3cy}) simplify to
\be\lab{3c}  m \ddot{x}+\gamma\dot{x}+\ka x=0,\ \  y=0.  \ee
Thus we obtain the classical damped oscillator equation from a
lagrangian theory at the expense of introducing an ``extra''
coordinate $y$, later constrained to vanish. Note that the
constraint $y(t)=0$ is {\it not} in violation of the equations of
motion since it is a true solution to Eqs.(\ref{3}) and
(\ref{3cy}).

We stress once more that the role of the ``doubled" $y$ coordinate
is absolutely crucial in the quantum regime since there it
accounts for the quantum noise in the fluctuating random force in
the system-environment coupling.

When one adopts the classical (legitimate) solution $y = 0$, the
$x$ system appears to be open, ``incomplete"; the knowledge of the
details of the processes inducing the dissipation may not always
be possible; these details may not be explicitly known and the
dissipation mechanisms are sometimes globally described by such
parameters as friction, resistance, viscosity etc.. In some sense,
such parameters are introduced in order to compensate the
information loss caused by dissipation. Such a loss of information
essentially amounts to neglecting the bath variables which
originate the damping and the fluctuations. Thus, by putting
$x_{+}=x_{-}$, i.e. by choosing $y = 0$, the quantum features are
washed out and one obtains the classical limit (see also
\cite{Blasone:1997xt}).

In quantum mechanics canonical commutation relations are not
preserved by time evolution due to damping terms. The role of
fluctuating forces is in fact the one of preserving the canonical
structure. According to our result, reverting from the classical
level to the quantum level, the loss of information occurring  at
the classical level due to dissipation manifests itself in terms
of ``quantum" noise effects arising from the imaginary part of the
action, to which the $y$ contribution is indeed crucial.

Going back to the dissipative quantum model, it can be shown
\cite{Celeghini:yv} that the Hamiltonian Eq. (\ref{1}), for each
given $\kappa$, can be obtained by the canonical quantization
procedure from the Lagrangian (\ref{2a}) with the choice Eq.
(\ref{P}). Note that the classical equations for the dho $x$ and
its time-reversal image $y$, Eqs. (\ref{3}) and (\ref{3cy}), are
associated in the canonical quantization procedure to the quantum
operators  $A$ and $\tilde A$. When we consider the quantum field
theory, the $A$ and the $\tilde A$ operators get labelled by the
(continuously varying) suffix $\kappa$ and for each $\kappa$ value
we have a couple of equations of the type (\ref{3}) and
(\ref{3cy}) for the field amplitudes \cite{Celeghini:yv}.

In the following sections we discuss some of the features of the
dissipative quantum model of brain related to the results
presented above.

\medskip

\section{Quantum noise induced entanglement}

We have seen that the doubled degrees of freedom account for the
quantum noise in the fluctuating random force in the
system-environment coupling. On the other hand, the doubled
degrees of freedom fully characterizes the structure of the space
of the states in the dissipative quantum model of brain. In other
words, the brain processes are intrinsically and inextricably
dependent on the quantum noise in the fluctuating random force in
the brain-environment coupling: there is a permanent
brain-environment entanglement.

The study of the brain-environment entanglement (namely the
entanglement between the ${A}_{\kappa}$ and the ${\tilde
A}_{\kappa}$ modes) is the purpose of the present section. To do
that we need to shortly summarize some aspects of the state space
of the dissipative quantum model.

Denote by $\{ | {\cal N}_{A_{\kappa}} , {\cal N}_{\tilde
A_{\kappa}}  \rangle \}$ the set of simultaneous eigenvectors of
${\hat N}_{A_{\kappa}} \equiv A^{\dagger}_{\kappa} A_{\kappa}$ and
${\hat N}_{\tilde A_{\kappa}} \equiv {\tilde A_{\kappa}}^{\dagger}
{\tilde A_{\kappa}}$, with ${\cal N}_{A_{\kappa}}$ and ${\cal
N}_{\tilde A_{\kappa}}$ non-negative integers and denote by
$|0\rangle_{0} \equiv | {\cal N}_{A_{\kappa}} = 0 , {\cal
N}_{\tilde A_{\kappa}} = 0 \rangle$ the state annihilated by
$A_{\kappa}$ and by ${\tilde A_{\kappa}}$: $A_{\kappa}
|0\rangle_{0} = 0 = {\tilde A_{\kappa}}|0\rangle_{0} $ for any
$\kappa$.

For definitiveness let us consider an initial time, say $t_{0} =
0$. The memory state is defined to be a zero energy eigenstate
(the vacuum) of $H_{0}$. The form of $H_{0}$ (cf. Eq. (\ref{2}))
then implies that the memory state is a condensate of {\it equal
number} of modes $A_{\kappa}$ and mirror modes ${\tilde
A}_{\kappa}$ for any $\kappa$. Thus, we may have infinitely many
memory states at $t_{0}$, each one corresponding to different
numbers ${\cal N}_{A_{\kappa}}$ of $A_{\kappa}$ modes, for all
${\kappa}$, provided  ${\cal N}_{A_{\kappa}} - {\cal N}_{{\tilde
A}_{\kappa}} = 0$ for all ${\kappa}$. We observe that the
commutativity of $H_{0}$  with ${H}_{I}$ ($[ H_{0} , H_{I} ]
 = 0$) ensures that the number $ ( {\cal N}_{A_{\kappa}} -
{\cal N}_{{\tilde A}_{\kappa}})$ is a constant of motion for any
$\kappa$. The $A_{\kappa}$ and ${\tilde A}_{\kappa}$ modes are
actually quasi-massless, i.e. they have a non-zero effective mass,
due to finite volume effects
\cite{Vitiello:1995wv,MyD,Alfinito:2001mm}.

Denote by ${|0\rangle }_{\cal N}$  the memory state with ${\cal N}
\equiv \{ {\cal N}_{A_{\kappa}} = {\cal N}_{{\tilde A}_{\kappa}},
\forall \kappa, at~~  t_{0} = 0 \}$ the set of integers defining
the "initial value" of the condensate, namely the code associated
to the information recorded at time $t_{0} = 0$.

The memory state is a two-mode ($SU(1,1)$ generalized) coherent
state (actually a two--mode squeezed state \cite{Vitiello:1995wv})
and is generated, at finite volume $V$, by the action of the
generator
\be\lab{G} G({\theta}) =- i \sum_{\kappa} \theta_{\kappa} \bigl (
A_{\kappa}^{\dagger} {\tilde A}_{\kappa}^{\dagger} -A_{\kappa}
{\tilde
 A}_{\kappa} \bigr )
~
 \ee
upon the state $|0\rangle_{0}$:
\be\lab{M} {|0 \rangle}_{\cal N} =
\exp{\bigl(-iG({\theta})\bigr)}|0\rangle_{0}
=\prod_k\;\frac{1}{\cosh\theta_{k}}\,\exp\left({-\tanh\theta_{k}
 A_k^{\dagger} {\tilde A}_{k}^{\dagger}}\right)\, |0\rangle_{0}
~. \ee

The average number ${\cal N}_{A_{\kappa}}$ is given by
\be\lab{num}
{\cal N}_{A_{\kappa}} =
 {_{\cal N}}\langle  0| A_{\kappa}^{\dagger} A_{\kappa}{|0\rangle}
 _{\cal N} = \sinh^{2} \theta_{\kappa}~,
\ee
which also relates the $\cal N$-set, ${\cal N} \equiv \{ {\cal
N}_{A_{\kappa}} = {\cal N}_{{\tilde A}_{\kappa}}, \forall \kappa,
at~~ t_{0}=0 \}$ to the $\theta$-set, $\theta \equiv \{
{\theta}_{\kappa}, \forall \kappa, at~~  t_{0} = 0 \}$. We also
use the notation ${\cal N}_{A_{\kappa}}(\theta) \equiv {\cal
N}_{A_{\kappa}}$ and ${|0(\theta) \rangle} \equiv {|0
\rangle}_{\cal N}$. In general we may refer to ${\cal N}$ or,
alternatively and equivalently, to the corresponding $\theta$, or
vice versa.

We note that ${|0\rangle}_{\cal N}$ is normalized to $1$ for all
$\cal N$:
\be\lab{ort} {}_{\cal N}\langle 0 | 0\rangle_{\cal N} = 1  \quad
\forall \, {\cal N}~. \ee
The state spaces $\{{|0\rangle}_{\cal N} \}$ and $\{{|0
\rangle}_{\cal N'}\}$ (representations of the canonical
commutation relations ({\it CCR}) of the operators $A_{\kappa}$
and ${\tilde A}_{\kappa}$) are each other unitarily inequivalent
for different codes ${\cal N} \neq {\cal N'}$ in the infinite
volume limit:
\be\lab{ort1} {}_{\cal N}\langle 0 | 0 \rangle_{\cal N'}
\mapbelow{V \rightarrow \infty} 0 \quad \forall \, {\cal N} \neq
{\cal N'}~. \ee

The whole space of states thus includes infinitely many unitarily
inequivalent representations $\{{|0\rangle}_{\cal N} \}$, for all
$\cal N$'s, of the {\it CCR}'s. The freedom introduced by the
degeneracy among the vacua ${|0\rangle}_{\cal N}$, for all $\cal
N$, solves the problem of memory capacity. A huge number of
sequentially recorded memories may {\it coexist} without
destructive interference since infinitely many vacua
${|0\rangle}_{\cal N} $  are independently accessible. Recording
information of code $\cal N'$ does not necessarily produce
destruction of previously printed information of code ${\cal N}
\neq {\cal N'}$. In the non-dissipative case this could not
happen. We thus realize the crucial role played in the state space
structure by the doubled degrees of freedom, namely by the
permanently present quantum noise in the brain-environment random
coupling. Such a noise contribution, represented by the mirror
modes, allows the possibility of introducing the $\cal N$-coded
``replicas" of the ground state, thus introducing a huge memory
capacity.

The quantum noise, which remarkably manifests itself in the
mentioned coherent structure of ${|0\rangle}_{\cal N}$, is also
responsible for its entangled nature. This goes as follows.

A two-mode state is an entangled state when it cannot be
factorized into two single-mode states. Inspection of Eq.
(\ref{M}) shows that ${|0\rangle}_{\cal N}$ can be written as
\be\label{M1}
  {|0\rangle}_{\cal N} = \left ( \prod_k\;\frac{1}{\cosh\theta_{k}} \right)\,\left (
|0 \rangle_{0} \otimes |{\tilde 0} \rangle_{0} - \sum_k
\;\tanh\theta_{k}
  \left( | A_k  \rangle \otimes |{\tilde A}_{k} \rangle
  \right)  + \dots \right)~.
\ee
Here the tensor product between the tilde and non-tilde sectors
has been explicitly expressed. Dots stand for higher power terms.
It is clear that the second factor in the r.h.s. of the above
equation cannot be reduced to the product of two single-mode
components and therefore $|0\rangle_{\cal N}$ can never be
factorized into two single-mode states. The state $|0
\rangle_{\cal N} $  may be also written as \cite{Celeghini:yv}:
\be\lab{M3}
  |0 \rangle_{\cal N} = \sum_{n=0}^{+\infty} \sqrt{W_n} \left( |n \rangle
  \otimes |{\tilde n} \rangle  \right)~,
\ee
\be\label{M4}
  W_n = \prod_k
  \frac{\sinh^{2n_k}\theta_{k}}{\cosh^{2(n_k+1)}\theta_{k}}\,,
\ee
where $n$ and $\tilde n$ denote the sets $\{ {n}_{\kappa} \}$ and
$\{{\tilde n}_{\kappa} \}$, respectively, and with $ 0 < W_n < 1$
 and $\sum_{n=0}^{+\infty} W_n = 1$. We have
\be\lab{M5} {}_{\cal N} \langle 0 |S_{A}|0 \rangle_{\cal N} =
\sum_{n=0}^{+\infty} W_n ln W_n~, \ee
with:
\be\lab{MM} S_{A} \equiv - \sum_{\kappa} \Bigl \{
A_{\kappa}^{\dagger} A_{\kappa} \ln \sinh^{2}  {\theta}_{\kappa} -
A_{\kappa} A_{\kappa}^{\dagger} \ln \cosh^{2} {\theta}_{\kappa}
\Bigr \}~. \ee
and thus $S_{A}$ can be interpreted as the entropy operator
\cite{Vitiello:1995wv,umezawa,Celeghini:yv} (we might as well
introduce $S_{\tilde A}$ given by replacing $A_{\kappa}$ and
$A_{\kappa}^{\dagger}$ with ${\tilde A}_{\kappa}$ and ${\tilde
A}_{\kappa}^{\dagger}$, respectively, in (\ref{MM})). Eq.
(\ref{M3}) shows that it provides a measure of the degree of
entanglement.

We stress that the entanglement is realized in the infinite volume
limit. It is expressed by the unitary inequivalence relation with
the vacuum $|0 \rangle_{0} \equiv |0 \rangle_{0} \otimes |{\tilde
0} \rangle_{0}$:
\be\lab{ort2} {}_{\cal N}\langle 0 | 0 \rangle_{0} \mapbelow{V
\rightarrow \infty} 0 \quad \forall \, {\cal N} \neq {0}~, \ee
which is indeed verified only in the continuum $\kappa$ limit,
i.e. the infinite volume limit (see also Eq. (\ref{ort1})).  The
probability of having the component state $|n, {\tilde n} \rangle
$ in the memory state $|0 \rangle_{\cal N}$ is $W_n$.  Since $W_n$
is a decreasing monotonic function of $n$, the contribution of the
states $|n, {\tilde n} \rangle $ would be suppressed for large $n$
at finite volume. In such a case, the tensor product of the tilde
and non-tilde sectors would involve only a finite number of $|n,
{\tilde n} \rangle $ component states and $|0 \rangle_{\cal N}$ be
different from the entangled state: a unitary transformation could
disentangle the tilde and non-tilde sectors. Finite volume effects
may thus spoil the entanglement. However, this is not the case in
the infinite volume limit, where the summation extends to an
infinite number of $n$--components.

We also note that the entanglement is generated by $G(\theta)$,
through which the brain is coupled with the environment.  By
considering the results of the previous section, we see that it is
exactly the quantum noise in the random force, coupling the system
(the brain) to the environment, the responsible for the
entanglement. Dissipation is therefore the root of the
entanglement and the robustness of the latter is in the
non-unitary character of the former. A similar result has been
also found at work in quite different physical contexts
\cite{Iorio:2002rr}.

Finally, we recall that the memory code ${\cal N}$ is a
macroscopic observable in the sense that it is not affected by
quantum fluctuations (this is a general feature of systems in
condensed matter physics where spontaneous symmetry breakdown
occurs). The stability of the order parameter against quantum
fluctuations is a virtue of the {\it coherence} of the DWQ boson
condensation. The ``change of scale" (from microscopic to
macroscopic scale) is thus dynamically achieved through the boson
condensation mechanism. The memory state ${|0\rangle}_{\cal N}$
provides thus an example of ``macroscopic quantum state" (other
examples of macroscopic quantum states in condensed matter physics
are the crystal state, the ferromagnetic state, the supeconducting
state, etc.).

In conclusion, the "brain (ground) state" may be represented as
the collection (or the superposition) of the full set of entangled
memory states ${|0\rangle}_{\cal N}$, for all $\cal N$.

In ``the space of the representations of the {\it CCR}'', each
representation $\{{|0\rangle}_{\cal N}\}$ may be thought as a
``point'' labelled by a given $\cal N$-set (or $\theta$-set) and
points corresponding to different $\cal N$ (or $\theta$) sets are
distinct points (do not overlap, cf. Eq. (\ref{ort1})). In other
words,  $\cal N$ (or $\theta$) is a {\it good} code. We also refer
to the space of the representations as to the ``memory space".

Till now our discussion has been limited to a given time $t_{0}$.
In the following section we analyze the time evolution of the
memory states. We will see that trajectories in the memory space
are chaotic trajectories.

\section{Time evolution and chaotic trajectories in the memory space}

In this section our task is to show that trajectories over the
representations of the {\it CCR} in the memory space are chaotic
trajectories. In order to see this, we will show that the
requirements characterizing the chaotic behavior in non-linear
dynamics are verified. These requirements can be formulated in a
standard way \cite{hilborn} as follows:

i)~ the trajectories are bounded and each trajectory does not
intersect itself (trajectories are not periodic).

%in the sense of Eq. (\ref{nr}).

ii)~~there are no intersections between trajectories specified by
different initial conditions.

% ii)  the trajectory in the space of the representations of the
%(memory) state $|0(t)\rangle_{\cal N}$ {\it does not crosses itself} as
%time evolves (it is not a periodic trajectory).

% iii) trajectories specified by different initial conditions
% (${\cal N} \neq {\cal N'}$) {\it never cross each other}.

iii) trajectories of different initial conditions are
 diverging trajectories.

We are referring to trajectories in the space of the
representations and, as shown in
\cite{Manka:pn,DelGiudice:tc,Vitiello:2003me}, they are {\it
classical} trajectories.

%iv) the ``distance" between points of two trajectories of
%different initial conditions (${\cal N} \neq {\cal N'})$ increases
%as $\frac{1}{2}e^{t\sum 2\Gamma_{\kappa}}$.*****

At finite volume $V$, the time evolution of the memory state
${|0\rangle}_{\cal N}$ is given by \cite{Vitiello:1995wv}
%
%$$
%| 0(t) \rangle_{\cal N} = \exp{\left ( - i t
%{{H}\over{\hbar}}\right )} |0\rangle_{\cal N} = \exp{\left ( - i t
%{{H_{I} }\over{\hbar}}\right )}
% |0\rangle_{\cal N}   %\noeqno
$$
| 0(t) \rangle_{\cal N} = \exp{\left ( - i t
{{H}\over{\hbar}}\right )} |0\rangle_{\cal N}
$$
\be\lab{timev}
= \prod_{\kappa} {1\over{\cosh{(\Gamma_{\kappa} t
  - {\theta}_{\kappa} )}}} \exp{
\left ( \tanh {(\Gamma_{\kappa} t - {\theta}_{\kappa}  )}
A_k^{\dagger} {\tilde A}_{k}^{\dagger} \right )} |0\rangle_{0}~,
\ee
where the commutativity between $H_{I}$ and $G(\theta)$ has been
used. Again, $| 0(t) \rangle_{\cal N}$ is a $SU(1,1)$ generalized
coherent state. It is also an entangled state and, with the due
changes, the discussion of the previous section on the
entanglement also applies to it. For any $t$
\be\lab{nr} {}_{\cal N}\langle 0(t) | 0(t)\rangle_{\cal N}  = 1~,
\ee
and  (for $ {\int \! d^{3} \kappa \, \Gamma_{\kappa}}$ finite and
positive), in the infinite volume limit,
\be\lab{t} {}_{\cal N}\langle 0(t) | 0\rangle_{\cal N} \mapbelow{V
\rightarrow \infty} 0 \quad~~ \forall \,
 t \quad ,
\ee
\be\lab{tt} {}_{\cal N}\langle 0(t) | 0(t') \rangle_{\cal N}
\mapbelow{V \rightarrow \infty} 0 \quad~~ \forall \, t\, , t'
\quad , \quad t \neq t' \quad . \ee
These relations express the unitary inequivalence of the states
$|0(t)\rangle_{\cal N} $ (and of the associated Hilbert spaces
$\{| 0(t) \rangle_{\cal N} \}$) at different time values $t \neq
t'$ in the infinite volume limit: the non-unitarity of time
evolution implied by damping is consistently recovered in the
unitary inequivalence among the representations $\{| 0(t)
\rangle_{\cal N} \}$ at different $t$'s in the infinite volume
limit. The non-unitary, damped time evolution is indeed manifest
in the relation:
\be\lab{t0} \lim_{t\to \infty} {}_{\cal N}\langle 0(t) |
0\rangle_{\cal N}  \, \propto \lim_{t\to \infty}
 \exp{\left ( -t  \sum_{\kappa}  \Gamma_{\kappa}  \right )} = 0~,
\ee
which holds provided $ {\sum_{\kappa} \Gamma_{\kappa} > 0}$.

Time evolution of the memory state $|0\rangle_{\cal N}$ is thus
represented as the (continuous) transition through the
representations $\{| 0(t) \rangle_{\cal N} \}$ at different $t$'s,
namely by the ``trajectory" through the ``points" $\{| 0(t)
\rangle_{\cal N} \}$ in the space of the representations. The
trajectory ``initial condition'' at $t_{0} = 0$ is specified by
the $\cal N$-set. As already mentioned, this is a classical
trajectory \cite{Manka:pn,DelGiudice:tc,Vitiello:2003me}:
transition between unitarily inequivalent representations would be
strictly forbidden in a quantum dynamics.

We observe that the trajectories are {\it bounded} in the sense of
Eq. (\ref{nr}), which shows that the ``length" (the norm) of the
``position vectors" (the state vectors at time $t$) in the
representation space is finite (and equal to one) for each $t$. We
recall that  $SU(1,1)$ consists of all unimodular $2 \times 2$
matrices leaving invariant the Hermitian form $|z_{1}|^{2} -
|z_{2}|^{2}$. Eq. (\ref{nr}) rests on such an invariance.
Moreover, the set of points representing the coherent states
$|0(t)\rangle_{\cal N}$ for any $t$ can be shown to be isomorphic
to the union of circles of radius ${r_{\kappa}}^{2} =
\tanh^{2}(\Gamma_{\kappa}t - \theta_{\kappa})$ for any $\kappa$
\cite{Perelomov}.

Eqs. (\ref{t}) and (\ref{tt}) express the fact that the trajectory
of given $\cal N$ {\it does not crosses itself} as time evolves
(it is not a periodic trajectory): the ``points"
$|0(t)\rangle_{\cal N}$ and $|0(t')\rangle_{\cal N}$ through which
the trajectory goes, for any $t$ and $t'$, with $t \neq t'$, after
the initial time $t_{0} = 0$, never coincide. We thus conclude
that the requirement $i)$ is satisfied.

Eqs. (\ref{t}) and (\ref{tt}) also hold for ${\cal N} \neq {\cal
N'}$ in the infinite volume limit:
\be\lab{tn} {}_{\cal N}\langle 0(t) | 0\rangle_{\cal N'}
\mapbelow{V \rightarrow \infty} 0 \quad~~ \forall \,
 t \quad ,~~\forall \, {\cal N} \neq {\cal N'}
\ee
\be\lab{ttn} {}_{\cal N}\langle 0(t) | 0(t') \rangle_{\cal N'}
\mapbelow{V \rightarrow \infty} 0 \quad~~ \forall \, t\, , t'
\quad ,~~\forall \, {\cal N} \neq {\cal N'}~. \ee
Eqs. (\ref{tn}) and (\ref{ttn}) have been obtained by using the
fact that in the continuum limit, for given $t$ and $t'$ and for
${\cal N} \neq {\cal N'}$, $\cosh({\Gamma}_{\kappa}t -
{\theta}_{\kappa} + {\theta'}_{\kappa})$ and
$\cosh({\Gamma}_{\kappa}(t - t') - {\theta}_{\kappa} +
{\theta'}_{\kappa})$, respectively, are never identically equal to
$1$ {\it for all} $\kappa$. Eq. (\ref{ttn}) is true also for $t =
t'$ for any ${\cal N} \neq {\cal N'}$. Eqs. (\ref{tn}) and
(\ref{ttn}) thus tell us that {\it trajectories specified by
different initial conditions (${\cal N} \neq {\cal N'}$) never
cross each other}. Requirement ii) is thus satisfied.

%We remark that, in the infinite volume limit, due to
The property ii) implies that no {\it confusion} (interference)
arises among different memories, even as time evolves. In
realistic situations of finite volume, states with different codes
may have non--zero overlap (the inner products (\ref{tn}) and
(\ref{ttn}) are not zero). This means that some {\it association}
of memories becomes possible. In such a case, at a ``crossing"
point between two, or more than two, trajectories, one can switch
from one of these trajectories  to another one which there
crosses. This may be felt indeed as association of memories or as
``switching" from one information to another one. This reminds us
of the ``mental switch" occurring, for instance, during the
perception of ambiguous figures \cite{A}, and, in general, while
performing some perceptual and motor tasks \cite{B,C} as well as
while resorting to free associations in memory tasks \cite{D}.

At each $t$, the average number of modes of type $A_{\kappa}$ is
given by
\be\lab{Nt} {\cal N}_{A_{\kappa}}(\theta,t) \equiv {}_{\cal
N}\langle 0(t) | A_{\kappa}^{\dagger} A_{\kappa} | 0(t)
\rangle_{\cal N}  = \sinh^{2}\bigl ( \Gamma_{\kappa} t -
{\theta}_{\kappa} \bigr )~, \ee
and similarly for modes of type ${\tilde A}_{\kappa}$. It can been
shown that this number satisfies the Bose distribution. Eq.
(\ref{Nt}) is actually a statistical average and the memory state
$| 0(t) \rangle_{\cal N}$ is found to be a thermal state
\cite{Vitiello:1995wv}.

Let us now study how the ``distance" between trajectories in the
memory space behave as time evolves. Let us consider two
trajectories of different initial conditions, ${\cal N} \neq {\cal
N'}$ ($\theta \neq {\theta}'$). We want to compute the time
evolution of the difference between the two {\it codes}. At time
$t$, each component ${\cal N}_{A_{\kappa}}(t)$ of the code ${\cal
N} \equiv \{ {\cal N}_{A_{\kappa}} = {\cal N}_{{\tilde
A}_{\kappa}}, \forall \kappa, at~~ t_{0}=0 \}$ is given by the
expectation value in the memory state of the number operator
$A_{\kappa}^{\dagger} A_{\kappa}$.  We have:
$$
\Delta {\cal N}_{A_{\kappa}}(t) \equiv {\cal
N'}_{A_{\kappa}}(\theta',t) - {\cal N}_{A_{\kappa}}(t) =
$$
\be\lab{co1} = \sinh^{2}\bigl ( \Gamma_{\kappa} t -
{\theta}_{\kappa} + {\delta \theta} \bigr ) - \sinh^{2}\bigl (
\Gamma_{\kappa} t - {\theta}_{\kappa} \bigr ) \approx \sinh \bigl
( 2(\Gamma_{\kappa} t - {\theta}_{\kappa}) \bigr ){\delta
\theta_{\kappa}} ~, \ee
where ${\delta \theta_{\kappa}} \equiv {\theta}_{\kappa} -
{\theta}'_{\kappa}$ (which, in full generality, may be assumed to
be greater than zero), and the last equality holds for small
${\delta \theta_{\kappa}}$ (i.e. for a very small difference in
the initial conditions of the two memory states). The
time-derivative then gives
\be\lab{co3} \frac{\partial}{\partial t}\Delta {\cal
N}_{A_{\kappa}}(t) = 2 {\Gamma_{\kappa}} \cosh \bigl (
2(\Gamma_{\kappa} t - {\theta}_{\kappa})  \bigr ){\delta
\theta_{\kappa}} ~. \ee
From this we see that the difference between originally even
slightly different ${\cal N}_{A_{\kappa}}$'s grows as a function
of time.  For large enough $t$, the modulus of the difference
$\Delta {\cal N}_{A_{\kappa}}(t)$ and its time derivative diverge
as $\exp{(2\Gamma_{\kappa} t) }$, for all $\kappa$'s. The quantity
$ 2 \Gamma_{\kappa}$, for each $\kappa$, appears thus to play a
role similar to that of the Lyapunov exponent in chaos theory
\cite{hilborn}. In conclusion, we see that trajectories in the
memory space, differing by a small variation $\delta \theta$ in
the initial conditions, diverge exponentially as time evolves.
This may account for the high perceptive resolution in the
recognition of the perceptual inputs.

Eq. (\ref{co1}) also shows that the difference between
$\kappa$--components of the codes $\cal N$ and $\cal N'$ may
become zero at a given time $t_{\kappa} =
\frac{\theta_{\kappa}}{\Gamma_{\kappa}}$. However, the difference
between the codes $\cal N$ and $\cal N'$ does not necessarily
become zero. The codes are made up by a large number (infinite in
the continuum limit) of ${\cal N}_{A_{\kappa}}(\theta,t)$
components, and they are different even if a finite number of
their components are equal. On the contrary, for ${\delta
\theta_{\kappa}} \equiv {\theta}_{\kappa} - {\theta'}_{\kappa}$
very small, suppose that the time interval $\Delta t = \tau_{max}
-\tau_{min}$, with $\tau_{min}$ and $\tau_{max}$ the minimum and
the maximum, respectively, of $t_{\kappa} =
\frac{\theta_{\kappa}}{\Gamma_{\kappa}}$, for {\it all}
$\kappa$'s, be ``very small''. Then the codes are ``recognized''
to be ``almost'' equal in such a $\Delta t$.  Eq. (\ref{co1}) then
expresses the ``recognition'' (or recall) process and we see how
it is possible that ``slightly different'' ${\cal
N}_{A_{\kappa}}$--patterns (or codes) are ``identified''
(recognized to be the ``same code'' even if corresponding to
slightly different inputs). Roughly, $\Delta t$ may be taken as a
measure of the ``recognition time''.

Let us finally recall that (see \cite{Vitiello:1995wv})
$\sum_{\kappa} E_{\kappa} {\dot {\cal N}}_{A_{\kappa}} dt =
\frac{1}{\beta} dS_{A}$, where $E_{\kappa}$ is the energy of the
mode $A_{\kappa}$, $\beta = \frac{1}{k_{B}T}$, $k_{B}$ the
Boltzmann constant. $dS_{A}$ is the entropy variation associated
to the modes $A$ and ${\dot {\cal N}}_{A_{\kappa}}$ denotes the
time derivative of ${\cal N}_{A_{\kappa}}$. Eq. (\ref{co3}) leads
then to the relation between the differences in the variations of
the entropy and the divergence of trajectories of different
initial conditions:
\be\lab{co4} \Delta \sum_{\kappa} E_{\kappa}{\dot {\cal
N}}_{A_{\kappa}}(t) dt = \sum_{\kappa} 2 E_{\kappa}
{\Gamma_{\kappa}} \cosh \bigl ( 2 (\Gamma_{\kappa} t
-{\theta}_{\kappa})\bigr ){\delta \theta_{\kappa}} dt =
\frac{1}{\beta} \bigl ( dS'_{A} - dS_{A} \bigr )~. \ee
An interesting question is the one of the relation between this
last equation and the relation between the Lyapunov exponent and
the Kolmogorov-Sinai entropy in the chaos theory \cite{hilborn}.
We plan to study such a question in a future work.

In conclusion, also the requirement iii) is satisfied.

Trajectories in the representation space have thus chaotic
behavior in the infinite volume limit. This result may have a
connection with experimental observation of chaotic behavior in
neural aggregates of the olfactory system of laboratory animals
\cite{Free} (see also \cite{K1}-\cite{K5}). In the present paper,
however, we do not analyze further such a possible connection.

\section{Concluding remarks}

The formalism of quantum dissipation is based on the doubling of
the system degrees of freedom, and we have seen that the doubled
modes  ${\tilde A}_{\kappa}$ account for the quantum noise in the
fluctuating random force in the system-environment coupling.
Remarkably, such a noise manifests itself through the coherent
structure of the ground state ${|0\rangle}_{\cal N}$. The
entanglement  of the $A_{\kappa}$ modes with the ${\tilde
A}_{\kappa}$ modes is permanent in the infinite volume limit and
in such a limit the trajectories in the memory space are classical
chaotic trajectories.

We have seen that the density matrix and the Wigner function
formalism leads in a natural way to the doubling of the system
modes. It is interesting that the Wigner function also enters in
the analysis of the quantum aspects of chaotic neuron dynamics
\cite{arecchi}. On the other hand, the chaotic behavior of the
trajectories in the memory space suggests a possible connection,
which, however, deserves to be further studied, with laboratory
observations on the neuronal chaotic behavior \cite{Free}.

From (\ref{Nt}) we see that at at $t = \tau$, with $\tau$ the
largest of the values ${\tau_{\kappa}} \equiv {
{{\theta}_{\kappa}}\over{\Gamma_{\kappa}}}$, the memory state
$|0\rangle_{\cal N}$ is reduced (decayed) to the "empty" vacuum
$|0\rangle_{0}$: the information has been {\it forgotten}, the
${\cal N}$ code is decayed. The time $t = \tau$ can be taken as
the life-time of the memory of code $\cal N$. For details on the
life-time of the $\kappa$-modes see \cite{Alfinito:2000ck}.
Considering time--dependent frequency for the DWQ, modes with
higher momentum have been found to possess longer life--time.
Since the momentum is proportional to the reciprocal of the
distance over which the mode can propagate, this means that modes
with shorter range of propagation survive longer. On the contrary,
modes with longer range of propagation decay sooner.  This
mechanism may
produce the formation%
\index{domain formation} of ordered domains of finite different
sizes with different degree of stability: smaller domains would be
the more stable ones \cite{Alfinito:2000ck}.
%\cite{AV2}.
On the other hand, since any value of the momentum is in principle
allowed to the DWQ, we also see that a scaling law is present in
the domain formation (any domain size is possible in view of the
momentum/size relation).

The memory state thus evolves into the "empty" vacuum
$|0\rangle_{0}$ which acts as a sort of attractor state. However,
as $t$ gets larger than $\tau$ we have
\be\lab{tt0} \lim_{t\to \infty} {}_{\cal N}\langle 0(t) |
0\rangle_{0} \, \propto \lim_{t\to \infty}
 \exp{\left ( -t  \sum_{\kappa}  \Gamma_{\kappa}  \right )} = 0~.
\ee
This shows that the state $|0(t)\rangle_{\cal N}$ "diverges" away
from the attractor state $| 0\rangle_{0}$ with exponential law (we
always assume ${\sum_{\kappa} \Gamma_{\kappa} > 0}$).

In order to avoid to fall into such an attractor, i.e. in order to
not forget certain information, one needs to "restore" the ${\cal
N}$ code by "refreshing" the memory by {\it brushing up} the
subject (external stimuli maintained memory). One has to recover
the whole ${\cal N}$-set, if the whole code is ``corrupted", or
``pieces" of the memory associated to those ${\cal N}_{\kappa}$,
for certain $\kappa$'s, which have been lost at $t_{\kappa} = {
{{\theta}_{\kappa}}\over{\Gamma_{\kappa}}}$. Restoring the code is
a sort of ``updating the register" of the memories since it
amounts to reset the memory clock to the (updated) initial time
$t_{0}$. We also observe that even after the time $\tau$ is passed
by, the code $\cal N$ may be recovered provided $t$ is not much
larger than $\tau$, namely, as far as the approximation
$\cosh({\Gamma}_{\kappa}t - {\theta}_{\kappa}) \approx \exp{\left
( -t \sum_{\kappa} \Gamma_{\kappa}  \right )}$ does not hold, cf.
Eq. (\ref{tt0}).

We note that the mentioned commutativity between $H_{I}$ (actually
$H$) and $G(\theta)$ guaranties that evolution in time does not
affect the ``measurement" of the code $\cal N$, i.e. at each
instant of time $t$ during the time evolution, provided $t$ is
smaller than $\tau$, one may know the ``initial conditions" of the
trajectory under consideration.

We observe that Eq. (\ref{Nt}) may  be also rewritten as
\cite{Vitiello:1995wv}
$$
{\cal N}_{A_{\kappa}}(\theta,t) \equiv {}_{\cal N}\langle  0(t) |
A_{\kappa}^{\dagger} A_{\kappa} | 0(t) \rangle_{\cal N}
$$
\be\lab{Ntt}
 = e^{-\beta
E_{\kappa}} {}_{\cal N}\langle  0(t) | A_{\kappa}^{\dagger}{\tilde
A}_{\kappa}^{\dagger} | 0(t) \rangle_{\cal N} = e^{-\beta
E_{\kappa}} {}_{\cal N}\langle  0(t) | A_{\kappa}{\tilde
A}_{\kappa} | 0(t) \rangle_{\cal N}~, \ee
Eq. (\ref{Ntt}) expresses the average number of modes $A_{\kappa}$
in terms of the Boltzmann distribution $\exp{(-\beta E_{\kappa})}$
times a factor describing the process of creation or annihilation,
respectively, of one couple $A_{\kappa}{\tilde A}_{\kappa}$ in the
state $| 0(t) \rangle_{\cal N}$. The creation and the annihilation
of the couple of modes $A_{\kappa}{\tilde A}_{\kappa}$ are indeed
equivalent processes in the quasi-equilibrium approximation (the
limit of stationary free energy) where $\tanh^{2}(\Gamma_{\kappa}
t - \theta_{\kappa}) = \exp{(-\beta E_{\kappa})}$, see
\cite{Vitiello:1995wv}.

Eq. (\ref{Ntt}) shows that, in the quasi-equilibrium
approximation, the ``measure" of ${\cal N}_{A_{\kappa}}(\theta,t)$
(its expectation value in the memory state at time $t$) is
weighted by the Boltzmann factor, which, for given $\beta$, is
larger for smaller $E_{\kappa}$, i.e. for smaller $\kappa$ (and,
vice versa, is smaller for larger $\kappa$). Since smaller
$\kappa$ modes are the short lived ones, the probability of
reading out the memory code, i.e. of recalling, is larger for
short-lived memory (and, vice versa smaller for long-lived
memory). This is true also for the memory printing process, since
it corresponds to the creation in the memory state of the couples
of $A_{\kappa}{\tilde A}_{\kappa}$ modes. Eq. (\ref{Ntt}) thus
provides a possible description of the fact that short-lived
memories are easily stored and easily recalled (which is of
crucial importance in our relational life with the external
world), and that, on the contrary, long-lived memories may require
much more work to be stored, according indeed to our familiar
experience. An interesting question is the one concerning those
extremely stable memories controlling, e.g., vital functions.
These memories are ``protected" from being recalled: in general
they ``cannot be recalled" under the influence of any external
input. According to the discussion presented  above, it would seem
that they are associated with codes whose components are
characterized by very high values of $\kappa$, so that the
probability weighting the process of annihilation or creation of
the associated couples $A_{\kappa}{\tilde A}_{\kappa}$ in their
memory state is near to zero.

\bigskip
\bigskip

{\bf Acknowledgments}
\smallskip

We acknowledge MIUR, INFN and INFM for partial financial support.
One of the authors (G.V.) is grateful to Tito Arecchi, to Harald
Atmanspacher and to Walter Freeman for useful discussions. G.V.
also thanks Harald Atmanspacher for his invitation to the
International Workshop on Aspects of Mind--Matter Research, held
in Wildbad Kreuth, Bavaria, Germany, in June 2003, where some of
the results here presented were reported.

\end{document}